\documentclass{aastex}          
\usepackage{spr-astr-addons}    
\usepackage{url}
\usepackage{dcolumn}            
\usepackage{multirow}           
\usepackage{bm}                 
\newcommand{\cbr}[1]{#1}
\newcommand{\beq}{\begin{equation}}
\newcommand{\eeq}{\end{equation}}
\newcommand{\bea}{\begin{eqnarray}}
\newcommand{\eea}{\end{eqnarray}}
\newcommand{\e}[1]{\ensuremath{\!\times \!\!10^{#1}}}
\newcommand{\wt}[1]{\widetilde{#1}}
\newcommand{\wb}[1]{\overline{#1}}
\newcommand{\mvec}[1]{\textbf{\em #1}}
\newcommand{\mten}[1]{\mathbf{#1}}
\newcommand{\mbfg}[1]{\bm{#1}}

\newcommand{\delX}{\nabla_{\! \mvec{X}}}
\newcommand{\delten}{\nabla_{\! 10}}

\newcommand{\mrow}[3]{\multirow{#1}{#2}{#3}}
\newcommand{\mcol}[3]{\multicolumn{#1}{#2}{#3}}

\newcolumntype{d}[1]{D{.}{.}{#1}}

\begin{document}
%
\title{Power law relating 10.7 cm flux to sunspot number}

\shorttitle{Power law relating 10.7 cm flux to sunspot number}
\shortauthors{R. W. Johnson}

\author{Robert W. Johnson}
\affil{Alphawave Research, Atlanta, GA 30238, USA} 


\begin{abstract}
To investigate the relation between observations of the 10.7 cm flux and the international sunspot number so that a physical unit may be ascribed to historical records, both polynomial and power law models are developed giving the radio flux as a function of sunspot number and \emph{vice versa}.  Bayesian data analysis is used to estimate the model parameters and to discriminate between the models.  The effect on the parameter uncertainty and on the relative evidence of normalizing the measure of fit is investigated.  The power law giving flux as a function of sunspot number is found to be the most plausible model and may be used to estimate the radio flux from historical sunspot observations.
\end{abstract}



\section{Introduction}
\label{sec:intro}

That a relation exists between the 2800 MHz 10.7 cm solar radio flux observed by ground stations and the sunspot number as defined by Wolf has long been known~\citep{covington-63125,hathaway-357}.  The correspondence of the 10.7 cm flux with other indicators of solar activity as well as mechanisms for its origin are discussed by~\citet{tapping-127321}.  That solar magnetic activity correlates with various geophysical processes is now well established~\citep{labitz-803393,svens-8122,rwj:jgr02,rwj:astro01}, and the sunspot number provides our longest continuous record of its level.  Putting the sunspot number onto a footing with physical units is of intrinsic interest to the solar theorist.

Building a mathematical model to describe the relation between two quantities of physical interest is a popular pastime, and deciding whether to accept or reject a model based on a quality of fit parameter is often done.  However, the essential question is not ``how well does this model fit the data'' but rather ``how much better does this model fit the data relative to another model.''  If a single model is all that is available, its quality of fit is irrelevant, as no better idea has presented itself.  In Bayesian analysis~\citep{Sivia:1996}, it is the ratio of the integrated evidence evaluated at the parameters of best fit which determines the relative plausibility of the models under consideration.  After evaluating the best fitting parameters, we will compare their evidence ratios to determine the most plausible model consistent with the data.  The nonlinearity inherent in the definition of the Wolf index proves particularly hard to model.


\begin{figure}[th]
\includegraphics[width=\columnwidth]{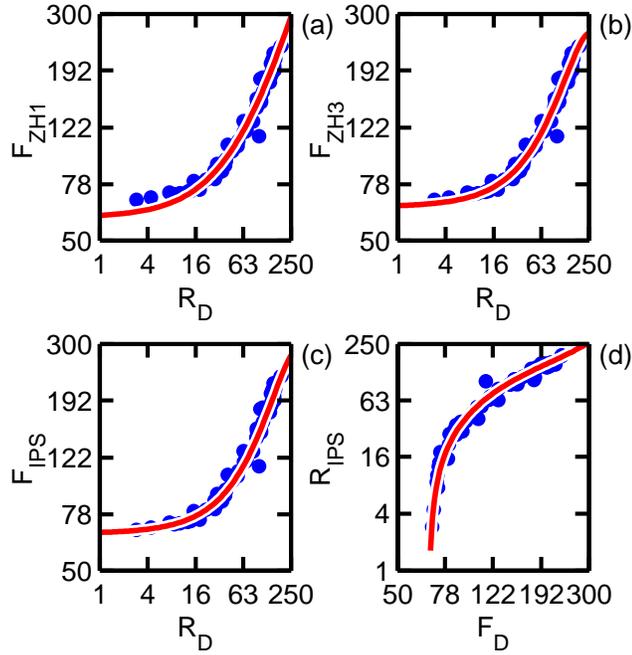}
\caption{Comparison of polynomial models available online using yearly data values} 
\label{fig:A}
\end{figure}

\begin{figure}[t]
\includegraphics[width=\columnwidth]{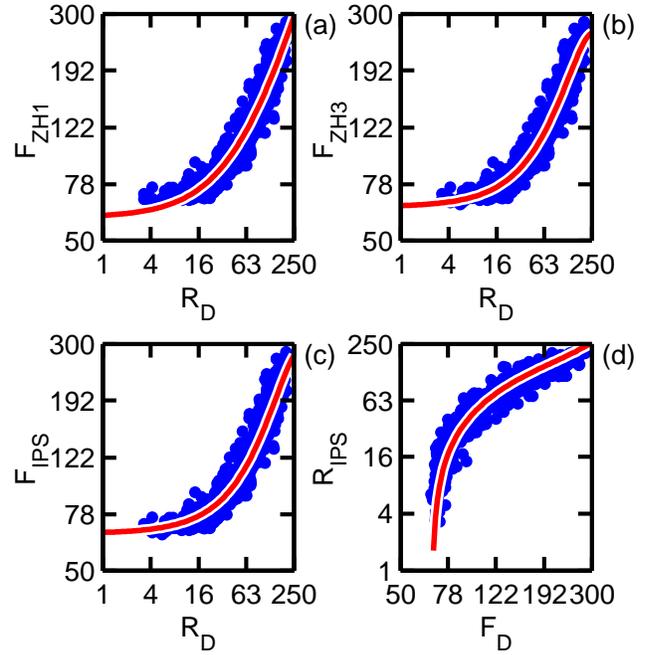}
\caption{Comparison of polynomial models available online using monthly data values} 
\label{fig:C}
\end{figure}

\section{Data selection and previous models}
\label{sec:data}

Often when comparing two independent sets of measurements, the choice of which data to use for abscissa and which for ordinate is not unambiguous.  Here we will consider polynomial and power law models each with three parameters relating the international sunspot number provided by the World Data Center for the Sunspot Index, Belgium~\citep{sidc:online}, to the adjusted Penticton/Ottawa 2800 MHz solar flux provided by the National Research Council of Canada and available through the National Geophysical Data Center, NOAA, USA.  The adjusted flux compensates for variation in the earth-sun distance.  These data sets do not quote variance values, which must then be set to unity for equal weighting of each data value.

\begin{table*}[t]
\small
\caption{Previous models available online: $F_{ZH1}$ and $F_{ZH3}$ are from~\citet{zhaohan-472} and $F_{IPS}$ and $R_{IPS}$ are from the IPS unit of the Australian Bureau of Meteorology} 
 \label{tab:A}
\begin{tabular}{r c l}
\tableline  
Model & &  Parameters  \\    
\tableline 
$F_{ZH1}$ & = & $60.1 + 0.932 \;R_D$                       \\
$F_{ZH3}$ & = & $65.2 + 0.633 \;R_D + 3.76\e{-3} \;R_D^2 - 1.28\e{-5} \;R_D^3$ \\
$F_{IPS}$ & = & $67.0 + 0.572 \;R_D + 3.31\e{-3} \;R_D^2 - 9.13\e{-6} \;R_D^3$\\
$R_{IPS}$ & = & $1.61 \;F_B - 5.37\e{-3} \;F_B^2 + 1.38\e{-5} \;F_B^3 \;,\; F_B = F_D - 67.0$ \\
\tableline 
\end{tabular}
\end{table*}

Previous investigators have usually selected a polynomial model for the relation between the solar flux $F_D$ and sunspot number $R_D$.  The subscript $D$ will be used to distinguish data values from model values.  \citet{zhaohan-472} consider both a linear fit and a cubic fit for $F(R_D)$ using annual values for 1947--2005, and the Ionospheric Prediction Service (IPS) unit of the Australian Bureau of Meteorology~\citep{IPS:FReqns} gives cubic equations for both $F(R_D)$ and $R(F_D)$ using monthly values from 1947--1990.  The radio flux is expressed in solar flux units (sfu) equal to $10^{-22} \mathrm{W/m^2/Hz}$.   These model equations, written in a form comparable to that which we will investigate, are displayed in Table~\ref{tab:A}.

A graphical comparison of these models using the yearly data values for 1947--2008 is given in Figure~\ref{fig:A}.  We see that the linear model is not capable of matching the data at low activity levels, and shortly beyond the region displayed the cubic models for $F(R_D)$ inflect downwards, implying a saturation of radio flux at extreme levels of solar magnetic activity.  The corresponding inverse relation $R_{IPS}(F_D)$ does not so inflect and is dominated by the cubic term at high flux levels.  These remarks hold as well for the monthly values shown in Figure~\ref{fig:C}.


\section{Bayesian data analysis}
\label{sec:bayes}

Our implementation of Bayesian data analysis draws primarily on the text by~\citet{Sivia:1996}.  The essential feature which takes it beyond simple regression is the use of a non-uniform prior in appropriate circumstances.  Using the language of conditional probabilities~\citep{Durrett:1994}, we write ``the probability of $A$ given $B$ under conditions $I$'' as \beq
{\rm prob}(A|B;I) \equiv p(A |_I B) \equiv p^A_B
\eeq when the background information $I$ is unchanging.  The choice of prior~\citep{dAgo:1998} represents one's background knowledge on the likely distribution of a parameter $x \in [x_0, x_1]$ before analysis of the current set of data.  A non-uniform prior $p^x_f$ arises naturally in many contexts, often representing a prior which is uniform over a change of variables $x \rightarrow F$ for some integrable function $f(x) = d F / d x$, with normalization $p^x_f = \Delta^{-1}_x f(x)$ for $\Delta_x \equiv \int_{x_0}^{x_1} f(x)\; d x$ such that $\int_{x_0}^{x_1} p^x_f\; d x = 1$.  Besides the uniform prior $f_x^{-1} = 1$, one commonly encounters the Jeffreys prior $f_x^{-1} = x$ uniform over $\log x$ and the Cauchy distribution $f_x^{-1} = 1 + x^2$ uniform over $\arctan x$.

\subsection{Parameter estimation}
\label{subsec:params}

One states Bayes' theorem in the context of parameter estimation as \beq \label{eqn:bayes}
p^\mvec{X}_D = p^\mvec{X} p^D_\mvec{X} / p^D \;,
\eeq reading ``the evidence for parameters $\mvec{X}$ given data $D$ equals the prior for $\mvec{X}$ times the likelihood for $D$ given $\mvec{X}$ divided by the chance of measuring $D$''.  What we call ``the evidence'' is often called ``the posterior'', as the normalization constant $p^D$ affecting neither parameter estimation nor model selection is sometimes called ``evidence''; both ``prior'' and ``likelihood'' have their usual meaning.  The logarithm (base $e$) of Equation~\ref{eqn:bayes} reads $L_E = L_P + L_L + \#_D$, where the final term is a constant equal to $- \log p^D$.  For independent data $\mvec{D} = \{D_t\}$ indexed by $t$ with Gaussian noise $\mbfg{\sigma}$, the likelihood factors as $p^\mvec{D}_\mvec{X} = \prod_t (2 \pi \sigma_t^2)^{-1/2} \exp (-\chi_t^2/2)$, where $\chi_t \equiv [M_t(\mvec{X}) - D_t]/\sigma_t$ is the weighted residual of the model $M$, so that $L_L$ has one term proportional to the measure of fit $\chi^2 \equiv \sum_t \chi_t^2$ and another which is constant.  With the definition of the merit function in terms of the model parameters, \beq
-L^\mvec{X} -L^\mvec{D}_\mvec{X} = - \sum_{x \in \mvec{X}} \log p^x + \dfrac{1}{2} \chi^2 + \#_{\mbfg{\sigma}} \;,
\eeq the problem becomes one of nonlinear global optimization~\citep{Press-1992}, seeking a unique solution to the equation $\delX L_E = 0$.  Short of evaluating the merit function over the entire prior range, one must rely on intuition and luck to varying degrees.  One's intuition, encoded in the form and domain of the prior functions $p^\mvec{X}$, contributes to the gradient of the log evidence $\delX L_E$ in the limit of poor data $\delX L_L \rightarrow 0$, thereby improving the chances of success.

Getting slightly ahead of ourselves, let us remark here that the traditional definition of the measure of fit $\chi^2$ is the {\em unnormalized } sum of weighted residuals squared.  \cbr{Recognizing that $\chi^2$ represents the variance of the data relative to the model,}  we believe that the normalized sum of weighted residuals is a more appropriate measure of fit, which one defines as \beq
\wt{\chi}^2 \equiv \sum_t \left[ M_t(\mvec{X}) - D_t \right]^2 d_t \;,
\eeq with the normalized weights $d_t \equiv  \sigma_t^{-2} / \sum_t \sigma_t^{-2}$ playing the role of the discrete measure factor.  For $N_t$ data values with unit variance $\sigma_t \equiv 1$, the normalized measure of fit reduces to $\wt{\chi}^2 = \chi^2 / N_t$.  In the continuum limit $\sum \rightarrow \int$ the measure factor is made apparent $\wt{\chi}^2 = \int [M(t) - D(t)]^2 d t$, and the normalization is required so that the measure of fit is not dependent upon the sampling rate---for an infinite or continuous data set, the unnormalized $\chi^2$ must be infinite for any model which does not perfectly match the data.  Replacing $M_t$ with a single parameter model given by the weighted mean of the data $\wb{D} = \sum_t D(t) (\sigma_t^{-2} / \sum_t \sigma_t^{-2})$ reveals the relationship between the measure of fit and the variance of the data vector $\wt{\chi}^2_{\wb{D}} = \sigma_\mvec{D}^2$.  The normalization has no effect on the location of the maximum likelihood solution $\mvec{X}_L$ while influencing the relative weighting of likelihood and prior in the expression for the evidence, thereby shifting the maximal evidence solution $\mvec{X}_E$ for non-uniform priors; it also affects the width of the error bars assigned to the parameter values, as $\exp(- \wt{\chi}^2/2) = [\exp(-\chi^2/2)]^{1/N_t}$.

\subsection{Model selection}
\label{subsec:modselec}

Given a single model, all one can do is estimate its best fitting parameters---the quality of fit is irrelevant beyond its role in the likelihood $p^\mvec{D}_\mvec{X}$ and its evidence may be normalized to unity.  However, faced with a choice of models, Bayes' theorem allows one to compute their evidence ratio $R_E^{AB}$, which reduces to the likelihood ratio \beq \label{eqn:evirat}
R_E^{AB} \equiv \dfrac{p^{A}_\mvec{D}}{p^{B}_\mvec{D}} = \dfrac{p^{A} p^\mvec{D}_{A} / p^\mvec{D}}{p^{B} p^\mvec{D}_{B} / p^\mvec{D}} \rightarrow \dfrac{p^\mvec{D}_A}{p^\mvec{D}_B} \equiv R_L^{AB} \;,
\eeq where $p^A = p^B$ indicates no prior preference for either model.  The null hypothesis of ``no relation'' is supported only when one can define a noise model, as the likelihood cannot be computed for a model $B$ given only that $M_t(B) \neq M_t(A)$.  The likelihood for each model is the unnormalized integral of the evidence for its parameters, \beq
p^\mvec{D}_{M} = \int p^{\mvec{D},\mvec{X}}_{M} d \mvec{X} = \int p^\mvec{X}_{M} p^\mvec{D}_{\mvec{X},M} d \mvec{X} \;,
\eeq and may be identified as the ``integrated probability bump'' over the model parameters $\mvec{X}$.  There is an unfortunate confusion of nomenclature in the literature because $p^\mvec{D}_M$ appears both in the position of chance in Equation~(\ref{eqn:bayes}) and in the position of likelihood in Equation~(\ref{eqn:evirat}).

Under the quadratic approximation, generally acceptable when the evidence is not severely truncated by the prior range, one can evaluate the integral analytically to write the negative logarithm of the likelihood as \beq \label{eqn:negloglike}
-L^\mvec{D}_M \approx \dfrac{1}{2} \chi^2 + \sum_k \log f^{-1}_k - \sum_k \log \left( \dfrac{\sqrt{2 \pi / h_k}}{\Delta_k} \right) \;,
\eeq for $\mvec{X}$ indexed by $k$ and $\{h_k\}$ the eigenvalues of the inverse of the variance matrix for the parameters $\prod_k h_k = \mathrm{det}\, \mten{\Sigma}_{\mvec{X}}^{-1}$, where the first two terms are the value of the merit function evaluated at its minimum and the remainder comprise the Occam factor accounting for the ratio of the width of the evidence $\mten{\Sigma}_{\mvec{X}}$ to the prior volume $\{\Delta_k\}$.  An additional parameter must provide not just a better fit but a significantly better fit in order for its plausibility to increase.  With several models to choose from, the one with the lowest value of $-L^D_M$ is deemed the most plausible, with the preference factor given by the exponential of the difference between the (negative) log evidence for each. 


\begin{table}[t]
\small
\caption{Summary of \cbr{prior functions $f_x$ and domains $[x_0,x_1]$ for the various models $M$}} 
 \label{tab:B}
\begin{tabular}{c|cc|cc|cc}
\tableline  
\mrow{2}{*}{$M$} & \mrow{2}{*}{$x$} & \mrow{2}{*}{$f_x^{-1}$} & \mcol{2}{c|}{Yearly} & \mcol{2}{c}{Monthly}  \\    
 & & & $x_0$ & $x_1$ & $x_0$ & $x_1$ \\
\tableline 
\mrow{3}{*}{$F_1$}
 & $B$ & 1 & 55 & 75 & 55 & 75 \\
 & $A$ & $1+A^2$ & 0 & 2 & 0 & 2 \\
 & $C$ & $1+C^2$ & -0.01 & 0.01 & -0.01 & 0.01 \\
\tableline 
\mrow{3}{*}{$F_2$}
 & $B$ & 1 & 55 & 75 & 55 & 75 \\
 & $A$ & $A$ & 0.05 & 5 & 0.05 & 5 \\
 & $C$ & $C$ & 0.05 & 5 & 0.05 & 5 \\
\tableline 
\mrow{3}{*}{$R_1$}
 & $B$ & 1 & 55 & 75 & 55 & 75 \\
 & $A$ & $1+A^2$ & 0 & 2 & 0 & 2 \\
 & $C$ & $1+C^2$ & -0.01 & 0.01 & -0.01 & 0.01 \\
\tableline 
\mrow{3}{*}{$R_2$}
 & $B$ & 1 & 55 & 69 & 55 & 66.5 \\
 & $A$ & $A$ & 0.05 & 5 & 0.05 & 5 \\
 & $C$ & $C$ & 0.05 & 5 & 0.05 & 5 \\
\tableline 
\end{tabular}
\end{table}

\section{Evaluation of the models}
\label{sec:modeval}

With two functional forms, polynomial and power law, and an arbitrariness to the selection of abscissa and ordinate, we consider a total of four models, two for $F(R_D)$ and two for $R(F_D)$.  As the 10.7 cm flux is observed never to fall below some background level $\sim$65 sfu, we use parameter $B$ for the background level in all models.  Parameter $A$ will be an amplitude, and parameter $C$ will be either another amplitude or the exponent in the power law.  Specifically, we consider the three parameter models given by \bea
F_1 &=& B + A R_D + C R_D^2 \;, \\
F_2 &=& B + (A R_D)^C \;, \\
R_1 &=& A (F_D - B) + C (F_D - B)^2 \;, \\
R_2 &=& A^{-1} (F_D - B)^{1/C} \;,
\eea where $F_D$ and $R_D$ are the data selected for the abscissa and the form of $R_2$ is chosen to compare directly its parameters with those of $F_2$.  We will be neglecting any influence from a lag between the solar flux and sunspot numbers~\citep{wilson-111279,sparavigna-2008}.  \cbr{Upon a visual inspection of the normalized monthly data series, any lag appears to be negligible at that temporal resolution.}

We summarize our use of priors in Table~\ref{tab:B}.  A uniform prior is assigned to $B$ whose \cbr{domain is adjusted for model $R_2$, which} requires $B \leq \min \{F_D\}$.  The Cauchy distribution serves as the prior for the amplitudes of the polynomial models, and for the power law models the Jeffreys prior is taken for $A$ and $C$.  The Jeffreys and Cauchy priors share the property that they may be used equally for the forward and inverse models of $F_2$ and $R_2$.  As $p^{1/A} \lvert d A^{-1} / d A \rvert = p^A \propto A^{-1}$, one may substitute $\tilde{A} = A^{-1}$ to write $p^{\tilde{A}} \propto \tilde{A}^{-1}$, and similarly for the Cauchy prior.  \cbr{Our results are not influenced greatly by the choice of priors, indicating that the fit is driven primarily by the likelihood.}

\begin{table}[t]
\small
\caption{\cbr{Yearly analysis results for best fitting parameters $B$, $A$, and $C$, their standard deviations $\sigma$, the quality of fit $\delten L_E$ and $-L_E$, and the evidence ratio $R_E^{21}$, for both the unnormalized and normalized log likelihood $-L_L$}} 
 \label{tab:C}
\begin{tabular}{c|c|d{2}d{3}d{4}|d{2}|c}
\tableline  
\mrow{2}{*}{$-L_L$} & \mrow{2}{*}{$M$} & \mcol{1}{c}{$B$} & \mcol{1}{c}{$A$} & \mcol{1}{c|}{$C$} & \mcol{1}{c|}{$\delten L_E$} & \mrow{2}{*}{$R_E^{21}$} \\    
 & & \mcol{1}{c}{$\sigma_B$} & \mcol{1}{c}{$\sigma_A$} & \mcol{1}{c|}{$\sigma_C$} & \mcol{1}{c|}{$-L_E$} & \\ 
\tableline 
\mrow{4}{*}{$\dfrac{\chi^2}{2}$}
 & \mrow{2}{*}{$F_1$}
  &  62.87 &   0.835 &   0.0005 &  -9.02  \\
& &   0.31 &   0.009 &   0.0001 &  \mcol{1}{r|}{2216} & $9.69$ \\
\cline{2-6}
 & \mrow{2}{*}{$F_2$}
  &  64.98 &   0.582 &   1.0970 &  -10.32 & $\! \! \times \! 10^{11}$ \\
& &   0.39 &   0.021 &   0.0082 &  \mcol{1}{r|}{2189} \\
\tableline 
\mrow{4}{*}{$\dfrac{\wt{\chi}^2}{2}$}
 & \mrow{2}{*}{$F_1$}
  &  63.01 &   0.830 &   0.0005 & -10.62 & \mrow{4}{*}{4.42} \\
& &   2.46 &   0.072 &   0.0004 &  44.23 \\
\cline{2-6}
 & \mrow{2}{*}{$F_2$}
  &  65.54 &   0.550 &   1.1105 & -10.84 \\
& &   3.02 &   0.148 &   0.0640 &  42.74 \\
\tableline 
\mrow{4}{*}{$\dfrac{\chi^2}{2}$}
 & \mrow{2}{*}{$R_1$}
  &  62.75 &   1.245 &  -0.0012 &  -8.89 \\
& &   0.25 &   0.010 &   0.0001 &   \mcol{1}{r|}{2403} & $2.04$ \\
\cline{2-6}
 & \mrow{2}{*}{$R_2$}
  &  67.01 &   0.402 &   1.1979 &  -10.40 & $\! \! \times \! 10^{36}$ \\
& &   0.27 &   0.011 &   0.0081 &   \mcol{1}{r|}{2319} \\
\tableline 
\mrow{4}{*}{$\dfrac{\wt{\chi}^2}{2}$}
 & \mrow{2}{*}{$R_1$}
  &  62.63 &   1.239 &  -0.0012 & -10.47 & \mrow{4}{*}{13.2} \\
& &   1.98 &   0.077 &   0.0005 &  47.59 \\
\cline{2-6}
 & \mrow{2}{*}{$R_2$}
  &  67.30 &   0.388 &   1.2082 & -11.07 \\
& &   2.04 &   0.083 &   0.0621 &  45.00 \\
\tableline 
\end{tabular}
\end{table}

\subsection{Yearly analysis}
The results for our analysis of the yearly data values are presented in Table~\ref{tab:C}, where the logarithm of the norm of the gradient at the solution $\mvec{X}_E$ is headed by $\delten L_E \equiv \log_{10} \lvert \delX L_E(\mvec{X}_E) \rvert$ and the negative log of the integrated evidence by $-L_E$.  The normalization of the measure of fit is indicated in the first column, and the evidence ratio $R_E^{21}$ is in the last column.  As $B(\wt{\chi}^2,R_2)$ is within $3 \sigma_B$ of its upper limit, a numerical evaluation of its integrated evidence is necessary, which differs from the approximate value by only a few percent.

\begin{figure}[t]
\includegraphics[width=\columnwidth]{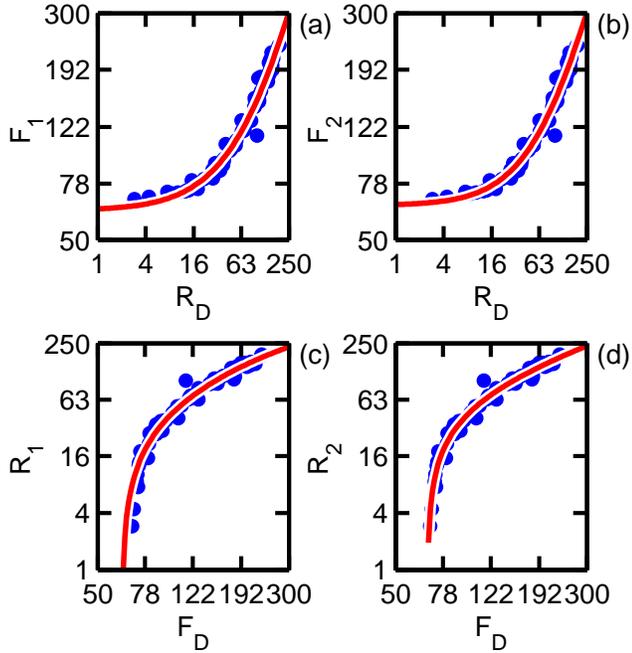}
\caption{Comparison of the best fitting solutions to the yearly data values using the normalized measure of fit $\wt{\chi}^2$} 
\label{fig:B}
\end{figure}

We see that the various models give slightly different estimates for the background radio flux $B$.  The polynomial models $F_1$ and $R_1$ return a value of about 63 sfu, while the power law model values are higher, around 65 sfu for $F_2$ and 67 sfu for $R_2$.  These remarks hold for either normalization of the measure of fit.  We compare in Figure~\ref{fig:B} the model solutions using the normalized measure of fit for all four models---the solutions for the unnormalized measure of fit are visually indistinguishable.  Compared to Figure~\ref{fig:A}, one can see that the power law $F_2$ provides with three parameters a quality of fit on par with a polynomial of four parameters and does not suffer from inflection problems at high levels of solar activity.  Polynomial models are notorious for having difficulties with extrapolation.

While the solution location $\mvec{X}_E$ is not greatly influenced by the choice of $\chi^2$ or $\wt{\chi}^2$ in $L_L$, the width of the marginal error bars is greater when using the normalized variance.  This change in the width of the evidence has a strong impact on the evaluation of its integral through the Occam factor in Equation~(\ref{eqn:negloglike}).  Consequently, the evidence ratio $R_E^{21}$ indicating the preference factor for the power law over the polynomial model is vastly different for the two choices of $L_L$.  With such similarity in the model solutions $\mvec{X}_E(\chi^2)$ and $\mvec{X}_E(\wt{\chi}^2)$, it is hard for us to countenance a preference factor on the order of $10^{36}$ or even $10^{12}$.  Using the normalized model variance $\wt{\chi}^2$ gives a preference factor $\sim\! 10$ for the power law models.  Furthermore, it seems reasonable to expect the variance of the background estimate $B$ for each model to be on the order of the variance between the models, as is found when using $\wt{\chi}^2$.

\subsection{Monthly analysis}

\begin{figure}[t]
\includegraphics[width=\columnwidth]{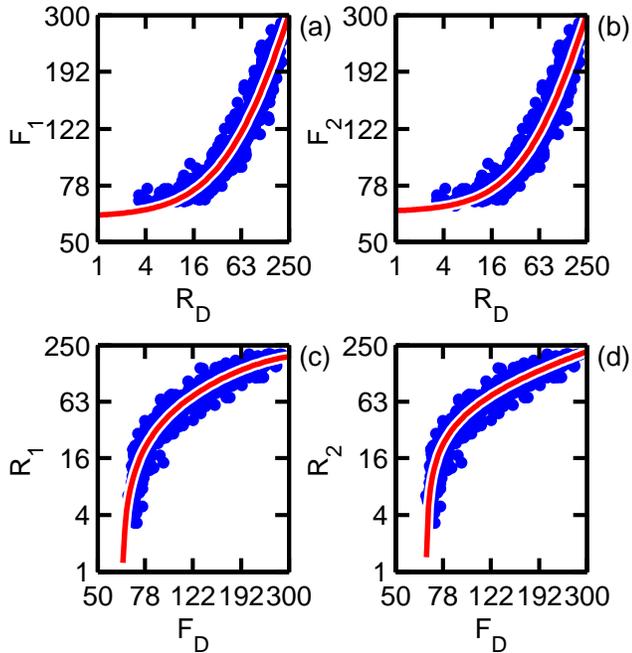}
\caption{Comparison of the best fitting solutions to the monthly data values using the normalized measure of fit $\wt{\chi}^2$} 
\label{fig:D}
\end{figure}

Repeating the analysis using the monthly data values, we find the results shown in Table~\ref{tab:D}.  While the assessment of the models for $F(R_D)$ is consistent with that of the yearly data, here we find that the power law model for $R(F_D)$ is not to be preferred.  The reason is because the background parameter $B$ is very tightly constrained to a value just below the minimum of the abscissa data $F_D$.  One might consider a modification of the model so that $R_2(F_D<B)=0$; however, such approach poses difficulties with the analytic evaluation of the gradient of the log likelihood.  The estimates of the background for the models $F(R_D)$ are lower compared to those from the yearly data, while those for $R_1$ are about the same, as are the remainder of the parameters.  We display the model solutions for the monthly data in Figure~\ref{fig:D}.

\begin{table}[t]
\small
\caption{\cbr{Monthly analysis results to compare with Table~\ref{tab:C}}} 
 \label{tab:D}
\begin{tabular}{c|c|d{2}d{3}d{4}|d{2}|c}
\tableline  
\mrow{2}{*}{$-L_L$} & \mrow{2}{*}{$M$} & \mcol{1}{c}{$B$} & \mcol{1}{c}{$A$} & \mcol{1}{c|}{$C$} & \mcol{1}{c|}{$\delten L_E$} & \mrow{2}{*}{$R_E^{21}$} \\    
 & & \mcol{1}{c}{$\sigma_B$} & \mcol{1}{c}{$\sigma_A$} & \mcol{1}{c|}{$\sigma_C$} & \mcol{1}{c|}{$-L_E$} & \\ 
\tableline 
\mrow{4}{*}{$\dfrac{\chi^2}{2}$}
 & \mrow{2}{*}{$F_1$}
  &  60.72 &   0.900 &   0.0002 &  -7.76  \\
& &   0.09 &   0.003 &   0.0000 &  \mcol{1}{r|}{92524} & $4.46$ \\
\cline{2-6}
 & \mrow{2}{*}{$F_2$}
  &  62.72 &   0.686 &   1.0642 &  -8.45 & $\! \! \times \! 10^{119}$ \\
& &   0.12 &   0.007 &   0.0023 &  \mcol{1}{r|}{92249} \\
\tableline 
\mrow{4}{*}{$\dfrac{\wt{\chi}^2}{2}$}
 & \mrow{2}{*}{$F_1$}
  &  60.87 &   0.895 &   0.0003 & -10.37 & \mrow{4}{*}{4.55} \\
& &   2.49 &   0.071 &   0.0004 & 134.49 \\
\cline{2-6}
 & \mrow{2}{*}{$F_2$}
  &  63.36 &   0.645 &   1.0780 & -11.71 \\
& &   3.19 &   0.178 &   0.0615 & 132.98 \\
\tableline 
\mrow{4}{*}{$\dfrac{\chi^2}{2}$}
 & \mrow{2}{*}{$R_1$}
  &  62.42 &   1.363 &  -0.0024 &  -7.88  & \mrow{4}{*}{$\sim 0$} \\
& &   0.06 &   0.002 &   0.0000 & \mcol{1}{r|}{80922} \\
\cline{2-6}
 & \mrow{2}{*}{$R_2$}
  &  66.50 &   0.281 &   1.3305 &  -7.98  \\
& &   0.00 &   0.001 &   0.0014 & \mcol{1}{r|}{82917} \\
\tableline 
\mrow{4}{*}{$\dfrac{\wt{\chi}^2}{2}$}
 & \mrow{2}{*}{$R_1$}
  &  62.35 &   1.360 &  -0.0024 & -10.39 \\
& &   1.67 &   0.061 &   0.0004 & 119.51 & $5.86$ \\
\cline{2-6}
 & \mrow{2}{*}{$R_2$}
  &  66.50 &   0.279 &   1.3332 &  -9.97 & $\! \! \times \! 10^{-3}$ \\
& &   0.01 &   0.028 &   0.0381 & 124.65 \\
\tableline 
\end{tabular}
\end{table}

As we are most interested in ascribing to the historical record of sunspot activity a physical unit based on the solar radio flux, the consistency in preference for $F_2$ to $F_1$ indicates the power law model function may be used for either yearly or monthly analysis.  Using $\psi$ and $\mu$ to indicate the yearly and monthly solutions, we are tempted to compare boxes of apples to apples by looking at the difference between $L_E(F_2^\psi)$ and $L_E(F_2^\mu)$, made possible through the use of the normalized measure of fit $\wt{\chi}^2$.  Reading the values from the tables, one can state that $F_2$ fits the yearly data better than the monthly data by a factor of about $\exp (133-42.7) \sim 10^{39}$.  Continuing the analogy to models for apples and bananas, one finds that $F_2^\psi$ fits better than $R_2^\psi$ by a factor $\exp (45-42.7) \sim 10$.


\section{Discussion and Conclusions}

The primary difficulty the models face is in relating the international sunspot number derived from the original Wolf index to the physical flux measurements of the S-component oscillation at small magnitudes.  It stems from the behavior of $R_D \equiv k_i (10 g + s)$ for small values of spot and group numbers $s$ and $g$.  The Wolf index has a jump from 0 to 11 for the first spot observed, and while the modern discontinuity is reduced slightly by the international reduction coefficient $k_i$, it still represents a significant source of nonlinearity.

An interesting feature of Bayesian model selection is that the log evidence, Equation~(\ref{eqn:negloglike}), contains factors for both the quality of fit and the error bars on the parameters given by the determinant of the inverse variance matrix.  The consequence is that for models with a similar measure of fit and prior volume, probability theory actually prefers the one with the larger error bars.  The reason is because a greater range of its parameter space yields a model consistent with the data.

Concluding, we have considered various models of three parameters relating the 10.7 cm solar radio flux to the international sunspot number.  The parameters found using maximal evidence are consistent with those given by other investigators.  Model selection using the evidence ratio indicates that the power law determining the solar flux from the sunspot number is most consistent with the yearly data values.  That model may be used to ascribe to the historical sunspot record a value in solar flux units.


%
\acknowledgments
Sunspot data provided by the SIDC-team, World Data Center for the Sunspot Index, Royal Observatory of Belgium, Monthly Report on the International Sunspot Number, online catalogue of the sunspot index, 1947--2008.
Penticton/Ottawa 2800 MHz Solar Flux data provided by the National Research Council of Canada and available through the National Geophysical Data Center, NOAA, Boulder, Colorado, USA .

\urlstyle{sf}

%

\end{document}